\documentclass[10pt, prl,aps,twocolumn,preprintnumbers,superscriptaddress,amssymb,amsmath]{revtex4-2}
\usepackage{graphicx}
\usepackage[colorlinks=true, pdfstartview=FitV, linkcolor=red, citecolor=blue, urlcolor=blue]{hyperref}
\usepackage{amssymb}
\usepackage{amsmath}
\usepackage{bm}
\usepackage{etoolbox}
\AfterEndEnvironment{subequations}{\par\noindent\ignorespaces}

\newcommand{\rmi}{\mathrm{i}}
\newcommand{\rme}{\mathrm{e}}
\newcommand{\rmd}{\mathrm{d}}
\newcommand{\bnabla}{\boldsymbol{\nabla}}
\newcommand{\ds}{\displaystyle}

\newcommand{\bp}{{\boldsymbol{p}}}
\newcommand{\bx}{{\boldsymbol{x}}}

\newcommand{\bE}{{\boldsymbol{E}}}

\newcommand{\bV}{{\boldsymbol{V}}}
\newcommand{\bsigma}{\boldsymbol{\sigma}}
\newcommand{\ba}{{\boldsymbol{a}}}
\newcommand{\bb}{{\boldsymbol{b}}}
\newcommand{\bv}{{\boldsymbol{v}}}
\newcommand{\bu}{{\boldsymbol{u}}}

\newcommand{\bj}{{\boldsymbol{j}}}

\newcommand\sect[1]{{\it #1}---}

\begin{document}

\title{Quantum Metric Corrections to Liouville's Theorem and Chiral Kinetic Theory}

\author{Kazuya~Mameda}
\affiliation{Department of Physics, Tokyo University of Science, Tokyo 162-8601, Japan}
\affiliation{RIKEN iTHEMS, RIKEN, Wako, Saitama 351-0198, Japan}

\author{Naoki~Yamamoto}
\affiliation{Department of Physics, Keio University, Yokohama 223-8522, Japan}

\preprint{RIKEN-iTHEMS-Report-25}

\begin{abstract}
Quasiparticles may possess not only Berry curvature but also a quantum metric in momentum space. 
We develop a canonical formalism for such quasiparticles based on the Dirac brackets, and demonstrate that quantum metric modifies the phase-space density of states at ${\cal O}(\hbar^2)$, leading to corrections to Liouville’s theorem, kinetic theory, and related physical quantities. In particular, we show that, in the presence of an inhomogeneous electric field, quantum metric induces corrections to the energy density and energy current.  
Applied to chiral fermions, this framework provides a nonlinear extension of chiral kinetic theory consistent with quantum field theory.
Our work paves the way to potential applications of the quantum metric in high-energy physics and astrophysics.
\end{abstract} 

\maketitle

\sect{Introduction}%
The geometry of wavefunctions in Hilbert space, known as quantum geometry, has become a fundamental concept in modern physics. 
One key component of quantum geometry is the Berry curvature~\cite{Berry}, which has played a central role not only in condensed matter physics~\cite{Xiao2010}, but also in high-energy and astrophysical contexts~\cite{Kamada:2022nyt}, particularly through the formulation of chiral kinetic theory~\cite{Son:2012wh,Stephanov:2012ki,Son:2012zy}. This framework describes topological quantum transport phenomena such as the chiral magnetic effect~\cite{Vilenkin:1980fu,Nielsen:1983rb,Fukushima:2008xe} and the chiral anomaly~\cite{Adler:1969gk,Bell:1969ts}.

The other essential component of quantum geometry is the quantum metric, defined as the Riemannian metric on the Hilbert space that measures the infinitesimal distance between the eigenstates of a Hamiltonian~\cite{Provost1980}. 
The quantum metric has recently attracted considerable attention in condensed matter physics owing to its impact on nonlinear responses; 
see, e.g., Refs.~\cite{Liu_2024,Gao2025,Verma,Jiang_2025,yu2025quantum} for recent reviews.
Previous applications of the quantum metric to transport phenomena have primarily focused on electric currents in solids.
Yet, it should be noted that a key property of the Berry curvature is the modification of the phase-space density of states and Liouville’s theorem~\cite{Xiao:2005qw,Duval:2005vn}, and, as a consequence, the very definitions of conserved charge and current as well as the formulation of kinetic equations~\cite{Son:2012wh,Son:2012zy}.
This knowledge naturally raises the fundamental question of whether the presence of the quantum metric may also lead to similar modifications.

The main aim of this Letter is to show that quantum metric generally modifies the phase-space density of states at ${\cal O}(\hbar^2)$, thereby yielding corrections to Liouville’s theorem, kinetic theory, and related physical quantities. 
For this purpose, we develop a canonical formalism for quasiparticles with quantum metric based on the Dirac brackets~\cite{Dirac1950,Dirac}. 
In particular, we show that the corrections to the energy density and energy current in the presence of an inhomogeneous electric field are governed by the quantum metric.
Our approach built upon the modified Liouville's theorem uniquely identifies the intrinsic $\mathcal{O}(\hbar^2)$ geometric contributions and definitively resolves the inconsistencies in preceding works, especially those concerning the intrinsic conductivity of the nonlinear anomalous Hall effect; see e.g., Refs.~\cite{Gao:2014,Kaplan:2024,Das}.

As a concrete example of a nontrivial quantum metric, we then apply our formulation to chiral fermions, which provides a nonlinear extension of chiral kinetic theory in terms of the quantum metric.
By comparing with the corresponding results from the Wigner function formalism based on quantum field theory (see, e.g., Ref.~\cite{Hidaka:2022dmn} for a review), we verify that nonlinear responses of chiral fermions originate geometrically from the quantum metric in the kinetic regime. This highlights the fundamental role of the quantum metric in nonlinear responses of Weyl and Dirac fermions, relevant to high-energy physics and astrophysics, including relativistic heavy-ion collisions, the early universe, neutron stars, and supernovae.

As a simple yet nontrivial demonstration, in this work we consider a static, but generally inhomogeneous, external electric field without a magnetic field. The charge of quasiparticles is absorbed into the scalar potential or electric field. 

\sect{Quantum metric and canonical formalism}%
Let us consider a quasiparticle in a theory with Berry curvature and quantum metric, described by the effective Lagrangian up to ${\cal O}(\hbar^2)$ \cite{Kolodrubetz2017} (see also Refs.~\cite{Ren:2025zei,Yoshida:2025tuo} for recent works):
\begin{equation}
\label{eq:L}
L = \bp\cdot\dot{\bx} - \epsilon
 	- \hbar\ba\cdot\dot{\bp}
	- \frac{\hbar^2}{2}\dot{p}_i\dot{p}_j G_{ij},
\end{equation}
where $\epsilon = \epsilon_{\bp} + \Phi$ with $\epsilon_{\bp}$ being the classical energy of the quasiparticle and $\Phi(\bx)$ the scalar potential.
Also, ${\bm a}(\bp)$ is the Berry connection and $G_{ij}({\bm p})$ is the energy-normalized quantum metric (also referred to as the Berry connection polarizability) in momentum space~\footnote{%
This is an effective Lagrangian for a specific band, with interband contributions encoded in $G_{ij}$ and $\ba$; see the derivation for chiral fermions below.}.
The equations of motion read
\begin{equation}
\label{eq:ELeq}
 \dot{x}_i 
= v_i + \hbar\epsilon_{ijk}E_jb_k - \hbar^2 (\dot{p}_j \dot{p}_k \Gamma_{jki} + G_{ij}\ddot{p}_j), \ \ \ 
\dot{p}_i= E_i,
\end{equation}
where $\bv={\bm \nabla}_{\bp}\epsilon_{\bp}$ is the velocity, $\bE = -\bnabla \Phi$ is the electric field, $\bb = \bnabla_\bp\times\ba$ is the Berry curvature, and
\begin{equation}
 \Gamma_{ijk} = \frac{1}{2}(\partial_i^p G_{jk} + \partial_j^p G_{ik} -\partial_k^p G_{ij})
\end{equation}
is the Christoffel symbol associated with $G_{ij}$.
The same equations as Eq.~\eqref{eq:ELeq} are obtained in Refs.~\cite{Jain2025,Fontana2025,Yoshida:2025tuo,Ren:2025zei} (see also Ref.~\cite{Mehraeen_2024}).
Here and below, the derivatives are denoted by $\partial_i = \partial/\partial x_i$ and $\partial_i^p = \partial/\partial p_i$.

One prototype example with such a nontrivial Berry curvature and quantum metric is a two-level system \cite{Kolodrubetz2017} or a Weyl Hamiltonian for chiral fermions [see Eqs.~(\ref{eq:Gij}) below]. 
The effective Lagrangian~\eqref{eq:L}, however, does not rely on the specific origin or the details of the quantum metric, and our argument based on it thus has broad applicability from condensed-matter systems to high-energy and astrophysical systems involving Weyl/Dirac fermions.
To make the present work self-contained and complete, we shall also provide a path-integral derivation of Eq.~\eqref{eq:L} for chiral fermions later in this Letter, extending the previous derivation of chiral kinetic theory in Ref.~\cite{Stephanov:2012ki}.

A characteristic of the Lagrangian with quantum metric in Eq.~\eqref{eq:L} is the presence of quadratic terms in $\dot p_i$, which are absent from the conventional theory up to ${\cal O}(\hbar)$ with the Berry connection.
Dynamical systems involving such terms are not directly suitable for the standard symplectic structure with Poisson brackets, which is usually the basis of the Liouville's theorem in phase space.
This difficulty can, however, be resolved by applying the Dirac-Bergmann theory of constrained systems~\cite{Dirac1950,Dirac,Bergmann1949}, which is essential, for instance, in the quantization of gauge theories~\cite{Faddeev:1969su}.
Namely, treating the original phase space as a constrained subspace on an extended whole space, one can obtain the symplectic structure of $\xi^a =(p_i,x_i)$ described by the Dirac brackets, a generalization of the Poisson brackets in constrained systems%
~\footnote{%
The usefulness of the Dirac bracket has already been recognized in the seminal work of the kinetic theory with the Berry curvature, Ref.~\cite{Sundaram-Niu}, though not explicitly investigated.}.

To this end, instead of $L$, we start from
\begin{equation}
 \tilde{L} = p_i\dot{x}_i 
    - \epsilon
 	- \hbar a_i q_i
	- \frac{\hbar^2}{2}G_{ij} q_iq_j 
    + \lambda_i (q_i-\dot{p_i}) .
\end{equation}
By treating $\lambda_i$ as Lagrange multipliers, the Euler-Lagrange equations for $p_i$, $x_i$, and $q_i$ are equivalent to Eqs.~\eqref{eq:ELeq}.
Let us define the generalized coordinates together with their conjugate momenta as
\begin{equation}
 \Xi^A = (p_i, x_i, q_i),
 \quad \Pi_A = (\pi^p_i,\pi^x_i, \pi^q_i) := \frac{\partial\tilde{L}}{\partial\dot{\Xi}^A},
\end{equation}
which constitute the 18-dimensional extended phase space $\Gamma$.
The Poisson bracket on $\Gamma$ is then defined as
\begin{equation}
 \{X,Y\}_\Gamma 
 = \frac{\partial X}{\partial \Pi_A}\frac{\partial Y}{\partial \Xi^A} - \frac{\partial X}{\partial \Xi^A}\frac{\partial Y}{\partial\Pi_A} .
\end{equation}
The Hamiltonian is obtained by the Legendre transformation as
\begin{equation}
\begin{split}
\label{eq:H_xpq}
 \tilde{H} = \Pi_A\dot\Xi^A-\tilde{L} 
 = \epsilon
 	+ (\pi^p_i + \hbar a_i)q_i
	+ \frac{\hbar^2}{2}G_{ij}q_iq_j  ,
\end{split}
\end{equation}
where we used $\pi^p_i = -\lambda_i$, $\pi^x_i = p_i$, and $\pi_i^q = 0$.

Although the original Lagrangian $L$ is described by the 12 independent variables $(p_i,x_i,\dot{p}_i,\dot{x}_i)$, the whole phase space $\Gamma$ itself is now 18-dimensional.
Therefore, we need to treat the present system as a constrained system.
The 12-dimensional subspace of $\Gamma$ is denoted by $\gamma$, where the coordinates and conjugate momenta are
$\xi^a = (p_i,x_i)$ and $\pi_a = (\pi^p_i,\pi^x_i)$, respectively.
The Poisson bracket in this subspace is defined as
\begin{equation}
 \{X,Y\}_\gamma 
 = \frac{\partial X}{\partial \pi_a}\frac{\partial Y}{\partial \xi^a} - \frac{\partial X}{\partial \xi^a}\frac{\partial Y}{\partial\pi_a} .
\end{equation}
From $\Pi_A = \partial\tilde{L}/\partial\dot{\Xi}^A$, the primary constraints read $\phi_i^x = \pi^x_i - p_i \approx 0$ and $\phi_i^q = \pi_i^q \approx 0$.
Here, ``$\approx$'' denotes weak equality, meaning that both sides are equated only after evaluating the Poisson brackets. Note that $\pi_i^p \approx - \lambda_i$ are not constraints but rather relations to determine the Lagrange multipliers $\lambda_i$, which do not belong to the set of phase-space variables.
The time evolution of the constraints is given by $\dot\phi^x_i = \{ \tilde{H},\phi^x_i\}_\Gamma$ and $\dot\phi^q_i = \{ \tilde{H},\phi^q_i\}_\Gamma$%
~\footnote{%
In the standard construction, the time evolution is described not by the original Hamiltonian $\tilde{H}$ but by the total Hamiltonian $\tilde{H}_\mathrm{T}:=\tilde{H}+u_i^x\phi_i^x+u_i^q\phi_i^q$ with Lagrange multipliers $u_i^x$ and $u_i^q$.
In the present case, however, one can show that the Poisson brackets between all the primary constraints vanish, %enabling us to describe 
so that the time evolution can be described in terms of $\tilde{H}$.}.
% $\dot\phi_i = \{ \tilde{H},\phi_i\}_\Gamma$.
To maintain consistency with the constraints above, we must impose $\dot{\phi}_i^x \approx 0$ and $\dot{\phi}_i^q \approx 0$, or equivalently, the following secondary constraints: $\varphi_i^x = \partial_i \epsilon + q_i \approx 0$ and $\varphi_i^q = \pi^p_i+\hbar a_i + \hbar^2 G_{ij}q_j \approx 0$.
After eliminating the degrees of freedom $(q_i,\pi_i^q)$ via the 6 constraints $\phi^q_i\approx 0$ and $\varphi_i^x \approx 0$, we obtain the canonical formalism described by
\begin{eqnarray}
&\label{eq:calH}
\displaystyle
\tilde{H}
    =\epsilon
 	- (\pi^p_i+\hbar a_i )\partial_i\epsilon
	+ \frac{\hbar^2}{2}\partial_i\epsilon \partial_j\epsilon G_{ij} ,\\
& \label{eq:phis}
\phi_{i} := \pi_i^x - p_i, \quad \varphi_i := \pi^p_i+\hbar a_i - \hbar^2 G_{ij}\partial_j\epsilon
\end{eqnarray}
with $\Phi_n := (\phi_i,\varphi_i)\approx 0$.

These 6 residual constraints properly restrict the dynamics in the 12-dimensional phase space $\gamma$ on the 6-dimensional subspace $g \ni \xi^a = (p_i,x_i)$.
The corresponding Dirac bracket is defined as
\begin{equation}
 \{X,Y\}_\mathrm{D} = \{X,Y\}_\gamma - \{X,\Phi_n\}_\gamma C^{nm}\{\Phi_m,Y\}_\gamma, 
\end{equation}
where $C^{nm}$ is the inverse of the constraint matrix $C_{nm}:=\{\Phi_n,\Phi_m\}_\gamma$ and is given by
\begin{equation}
\begin{split}
C^{nm}
&=
\begin{pmatrix}
  \hbar\epsilon_{ijk} b_k'
  & -\delta_{ij} -\hbar^2 G_{ik} \partial_j \partial_k \epsilon \\
  \delta_{ij} +\hbar^2 G_{jk} \partial_i \partial_k \epsilon
  & 0
\end{pmatrix},
\end{split}
\end{equation}
with
\begin{equation}
\label{eq:mod-ab}
 a_i' = a_i -\hbar G_{ij}\partial_j\epsilon,
 \quad
  b_i' =  \epsilon_{ijk}\partial_{j}^p a_k'.
\end{equation}
This $a_i'$ can be regarded as the modified Berry connection, and accordingly $b_i'$ as the modified Berry curvature~\cite{Gao:2014,Kaplan:2024}.

The full Hamiltonian (\ref{eq:calH}) under the constraints (\ref{eq:phis}) can be written as
\begin{equation}
\label{eq:H_tilde}
\tilde{H}
    \approx \epsilon
    - \frac{\hbar^2}{2} \partial_i \epsilon \partial_j \epsilon G_{ij}.
\end{equation}
This can be interpreted as the energy of the quasiparticle, including the correction from the quantum metric, which is consistent with the results of Refs.~\cite{Liu2022,Jia2024} under $\partial_i\epsilon = -E_i$.

\sect{Modified Liouville's theorem and kinetic theory}%
Now, the Lagrange bracket restricted on $g$, i.e., $\omega^{ab} = \{\xi^a,\xi^b\}_\mathrm{D}$ reads
\begin{equation}
\label{eq:LB}
 \omega^{ab} 
 = 
\begin{pmatrix}
  0 &  \delta_{ij} +\hbar^2 G_{ik} \partial_j \partial_k \epsilon \\
  -\delta_{ij} -\hbar^2 G_{jk} \partial_i \partial_k \epsilon
  & \hbar\epsilon_{ijk} b_k'
\end{pmatrix}.
\end{equation}
Up to ${\cal O}(\hbar^2)$, the determinant $\omega = \det(\omega_{ab})$ with $\omega_{ab}=(\omega^{ab})^{-1}$ is evaluated as
\begin{equation}
\begin{split}
\sqrt{\omega}
&= 1 - \hbar^2  G_{ij} \partial_i \partial_j\epsilon .
\end{split}
\end{equation}
For $\epsilon = \epsilon_{\bp} + \Phi$, the invariant measure of the phase space on $g$ is
\begin{equation}
\label{eq:measure}
 \rmd g = \sqrt{\omega}\,\rmd \xi = (1+\hbar^2\partial_i E_jG_{ij})\frac{\rmd\bx\rmd\bp}{(2\pi)^3}.
\end{equation}
This is a nonlinear extension of Refs.~\cite{Xiao:2005qw,Duval:2005vn} due to the quantum metric at ${\cal O}(\hbar^2)$, and constitutes one of our main results. 
Using the Lagrange bracket~\eqref{eq:LB}, we confirm that $\dot\xi^a = \{\tilde{H},\xi^a\}_\mathrm{D}$ reproduces the equations of motion~\eqref{eq:ELeq}.

The invariant measure (\ref{eq:measure}) implies that the probability of finding a particle in a phase-space cell is given by $\rho=\sqrt{\omega} n_{\bp}$, where $n_{\bp}=n_{\bp}(\epsilon_\bp')$ is the distribution function, with $\epsilon_\bp'$ being the modified quasiparticle energy up to ${\cal O}(\hbar^2)$ [see Eq.~(\ref{eq:H_tilde})],
\begin{equation}\label{eq:dispersion}
 \epsilon_\bp' = \epsilon_\bp - \frac{\hbar^2}{2}E_i E_j G_{ij}.
\end{equation}
From the modified Liouville's theorem $\rmd \rho/\rmd t = \partial_t \rho + \{\tilde{H},\rho\}_\mathrm{D}= 0$, we obtain the following kinetic equation up to ${\cal O}(\hbar^2)$: 
\begin{equation}
\begin{split}
\label{eq:keq}
 \Bigl[ \partial_t + \left(\bv' + \hbar \bE \times \bb' + \hbar^2 \bV \right)\cdot\bnabla + \bE\cdot\bnabla_\bp \Bigr] n_\bp = 0 ,
\end{split}
\end{equation}
where $\bv'=\bnabla_\bp \epsilon_\bp'$ and $V_k = -G_{kj} v_i \partial_i E_j$.
Here and below, we ignore the collision term; 
this is a standard procedure for extracting the intrinsic contributions.
Note that the Christoffel symbol in Eqs.~\eqref{eq:ELeq} is recast as the quantum-metric corrections in $\bv'$ and $\bb'$.
One can also confirm that, owing to the quantum metric correction in $\epsilon_\bp'$, the kinetic equation~\eqref{eq:keq} is satisfied by the equilibrium distribution $n_\bp^{\rm eq}=n_\mathrm{F}(\epsilon_\bp'-\mu)$ under the condition $\bnabla\mu=\bE$, where $n_\mathrm{F}(x)=(\rme^{x/T}+1)^{-1}$ is the Fermi-Dirac distribution function, with $\mu$ being the chemical potential and $T$ the temperature.

\sect{Number and energy densities, currents, and conservation laws}%
We are now ready to define the particle number density $n$ and current $\bj$, as well as the energy density $\varepsilon$ and energy current ${\bm J}$, for a general distribution function $n_\bp=n_\bp(\epsilon_\bp')$, not limited to the equilibrium case, $n_\bp^{\rm eq}$.
Multiplying Eq.~(\ref{eq:keq}) by the factors $\sqrt{\omega}$ and $\sqrt{\omega}\epsilon_\bp$ (not $\sqrt{\omega}\epsilon_\bp'$), and performing the momentum integration, we obtain the charge conservation, $\partial_t n + \bnabla\cdot \bj = 0$, and the energy conservation, $\partial_t \varepsilon + \bnabla\cdot {\bm J} = \bE\cdot\bj$, respectively, where
\begin{equation}
  n = \int_\bp \sqrt{\omega} n_\bp,
  \quad
  \bj = \int_\bp (\bv' + \hbar \bE \times \bb' + \hbar^2 \bV ) \sqrt{\omega} n_\bp,
  \label{eq:nj}
\end{equation}
and
\begin{equation}
\begin{split}
 \varepsilon
 = \int_\bp\sqrt{\omega} \epsilon_\bp' n_\bp,\quad
 {\bm J}
 = \int_\bp (\bv' + \hbar \bE \times \bb' + \hbar^2 \bV ) \sqrt{\omega} \epsilon_\bp' n_\bp,
 \label{eq:EJ}
\end{split}
\end{equation}
with $\int_\bp = \int \rmd\bp/(2\pi)^3$.
 
At ${\cal O}(\hbar^2)$, these $n$, $\bj$, $\varepsilon$, and ${\bm J}$ contain the contributions of quantum metric,
\begin{subequations}
\begin{align}
\label{eq:n2}
 n^{(2)}
 &= \hbar^2\int_\bp
 \biggl[
 -\frac{1}{2} E_iE_j G_{ij}\frac{\partial}{\partial \epsilon_{\bp}}
 + \partial_i E_j  G_{ij} 
 \biggr]n_\bp^{(0)}
 \\
 j^{(2)}_{k}
 &= \hbar^2 
 \int_\bp
 \biggl[
 E_iE_j 
 \partial_{[k}^p G_{i]j}
 +\partial_i E_j
 v_{[k}G_{i]j}
 \biggr]n_\bp^{(0)},
\label{eq:j2}
\end{align}
\label{eq:nj2}
\end{subequations}
and
\begin{subequations}
\begin{align}
 \label{eq:E2}
 \varepsilon^{(2)}
 &= \hbar^2  \int_\bp 
 \biggl[
 -\frac{1}{2}E_iE_j G_{ij}
 \biggl(
    1 + \epsilon_\bp
 \frac{\partial}{\partial\epsilon_\bp}
  \biggr)
 +\partial_i E_jG_{ij} \epsilon_\bp 
 \biggr]
 n_{\bp}^{(0)},  \\
 J_k^{(2)} 
 &= 
 \hbar^2 \int_\bp 
 \biggl[
    E_iE_j
    \biggl\{
        -\frac{1}{2}v_k G_{ij}
        \biggl(
            1+\epsilon_\bp
            \frac{\partial }{\partial \epsilon_\bp}
        \biggr)
        +\partial^p_{[k} G_{i]j}
        \epsilon_{\bp}
    \biggr\} \nonumber \\
 &\qquad\qquad
    +\partial_i E_j 
    v_{[k} G_{i]j}
    \epsilon_{\bp}
  \biggr] n_\bp^{(0)},
 \label{eq:J2}
 \end{align}
 \label{eq:EJ2}
\end{subequations}
where $n_\bp^{(0)} = n_\bp (\epsilon_\bp)$ and $X_{[i}Y_{j]} := X_iY_j-X_jY_i$.
While the $E^2$-dependent corrections in $\bj$~\cite{Gao:2014} and part of the $\partial_i E_i$-dependent contributions to $\bj$~\cite{Gao2019,Lapa2019} are known in the literature, the complete explicit forms of Eqs.~\eqref{eq:nj2} and~\eqref{eq:EJ2}, in particular the energy density $\varepsilon$ and the energy current ${\bm J}$, have not been presented. %to the best of our knowledge.
Importantly, $\partial_i E_i$-dependent corrections receive the contributions from the factor $\sqrt{\omega}$ in the modified invariant phase-space measure.
Although several mutually inconsistent results for the $E^2$-dependent contribution to $\bj$ have been reported in the literature, our result in Eq.~\eqref{eq:j2}, which agrees with Refs.~\cite{Gao:2014,Qiang_2025}, is distinguished by the fact that its validity is guaranteed by the modified Liouville's theorem and the corresponding conservation law.

\sect{Nonlinear responses of chiral and Dirac fermions}%
While so far our argument and derivation are generic and independent of the details of the Berry curvature and quantum metric, from now on we consider chiral and Dirac fermions, keeping their potential applications to high-energy physics and astrophysics.
This step also shows that the above results are consistent with those derived from quantum field theory, validating our formulation.

We first focus on a right-handed fermion. In this case, the classical energy dispersion is $\epsilon_{\bp} = |\bp|$, and the Berry curvature and the (energy-normalized) quantum metric are given by 
\begin{equation}
\begin{split}
\label{eq:Gij}
b_i = \frac{\hat{p}_i}{2|\bp|^2}\,, \quad
G_{ij} 
 =\frac{1}{4|\bp|^3} (\delta_{ij}-\hat{p}_i\hat{p}_j),
\end{split}
\end{equation}
respectively, where $\hat{p}_i = p_i/|\bp|$; see later for the derivation.
For chiral fermions, the spin is always locked to the momentum, and the quantum state is specified by the momentum orientation $\hat \bp$. The quantum metric $G_{ij}$, which characterizes changes in the quantum state, is thus transverse to $\hat \bp$, as in Eqs.~(\ref{eq:Gij}).
Equations (\ref{eq:dispersion}) and (\ref{eq:keq}), together with Eq.~(\ref{eq:Gij}), provide the dispersion relation and kinetic equation for the nonlinear chiral kinetic theory in the presence of an inhomogeneous electric field,
which enriches the symplectic structure through the phase space modification in Eq.~\eqref{eq:measure}.
To the best of our knowledge, an explicit form of the kinetic equation valid for inhomogeneous electric fields has not been derived within the Wigner-function formalism.

Let us consider the case of local equilibrium with local fluid velocity $\bu$. 
Then, the equilibrium distribution function is given by the summation of the contributions of particles and antiparticles as $n_\bp^{(0)\mathrm{eq}} = n_\mathrm{F}(\epsilon_\bp-\bu\cdot\bp-\mu) - n_\mathrm{F}(\epsilon_\bp-\bu\cdot\bp+\mu)$.
Expanding Eqs.~(\ref{eq:nj2}) to first order in $\bu$ and performing momentum integration, we extract the following $E^2$-dependent corrections:
\begin{subequations}
\begin{align}
 n^{(EE)} &= -\frac{\hbar^2C}{24\pi^2} |\bE|^2,
 \\
 \bj^{(EE)} 
 &= \frac{\hbar^2 C}{12\pi^2} 
 \Bigl[
        -\bE(\bE\cdot\bu)
       +|\bE|^2\bu
 \Bigr], 
 \label{eq:j_EE_hydro}
\end{align}
\label{eq:nj_EE_hydro}
\end{subequations}
where we introduced the integral
\begin{equation}
\label{eq:Jnm}
C
 = \int_0^\infty \frac{\rmd p}{p} \frac{\rmd}{\rmd p}
 \Bigl[
  n_\mathrm{F}(p-\mu)
  - n_\mathrm{F}(p+\mu)
 \Bigr].
\end{equation}
Equations~\eqref{eq:nj_EE_hydro} reproduce the results in Refs.~\cite{Yang:2020mtz,Mameda:2023ueq,Yang:2024sfp} derived from the Wigner function formalism based on the quantum field theory~\footnote{
We note that the covariant current $\tilde{j}^\mu$, derived in Refs.~\cite{Yang:2020mtz,Mameda:2023ueq,Yang:2024sfp}, is defined in the inertial frame.
Hence, it is related to our results through $\tilde{j}^0 = n^{(EE)}$ and $\tilde{\bm j}=n^{(EE)}\bu + \bj^{(EE)}$.}, 
where the relation to the geometric structure of the quantum metric was not appreciated. Although the integral $C$ cannot, in general, be evaluated analytically, one finds that $C
\rightarrow -{1}/{\mu}$ in the zero-temperature limit $T \rightarrow 0$~\cite{Yang:2020mtz}.

Similarly, Eq.~(\ref{eq:E2}) for chiral fermions in the hydrodynamic regime yields the $\partial_iE_i$-dependent correction,
\begin{equation}
\label{eq:E_hydro}
 \varepsilon^{(\partial E)}
 = \frac{\hbar^2\mu}{12\pi^2}\bnabla\cdot\bE,
\end{equation}
which also reproduces the result obtained by the Wigner function formalism in Refs.~\cite{Mameda:2023ueq,Yang:2024sfp}.

From Eqs.~\eqref{eq:nj2} and~\eqref{eq:EJ2}, one might expect to derive formulas for $n^{(\partial E)}$, $\bj^{(\partial E)}$, $\varepsilon^{(EE)}$, and ${\bm J}^{(EE)}$. 
For chiral fermions, the momentum integrals in these expressions are dominated by contributions around the origin $|\bp|\sim 0$, where the kinetic description breaks down.
For this reason, we do not present them explicitly here.

The extension to Dirac fermions is straightforward. While the Berry curvatures of right- and left-handed fermions have opposite signs and their contributions cancel unless a chirality imbalance is present (see, e.g., Ref.~\cite{Kamada:2022nyt} for a review), the quantum-metric contributions appear with the same sign for both, yielding a nonvanishing sum even without such an imbalance. Consequently, the effects of the quantum metric manifest at ${\cal}O(\hbar^2)$ in generic relativistic many-body systems composed of Dirac fermions, not restricted to chiral matter.
For example, in quark matter with $N_{\rm c}$ colors and $N_{\rm f}$ flavors, the correction to the energy density induced by a spatially inhomogeneous electric field is given by $2 N_{\rm c} N_{\rm f}$ times the second term in Eq.~(\ref{eq:E2}).

\sect{Path-integral derivation of the Lagrangian with quantum metric}%
Let us now derive the effective Lagrangian (\ref{eq:L}) for a right-handed fermion by extending of the path-integral  derivation of chiral kinetic theory in Ref.~\cite{Stephanov:2012ki} up to ${\cal O}(\hbar^2)$. The Weyl Hamiltonian is given by $ H_\mathrm{W} = \bp\cdot\bsigma + \Phi$.
The transition amplitude from initial to final states, $\mathcal{A}=\langle f|\rme^{\rmi H_\mathrm{W}(t_\mathrm{f}-t_\mathrm{i})/\hbar}|i\rangle$, is written as the path integral with this Hamiltonian.
After the diagonalization by the unitary matrix $V_\bp =
(u_{\bp+} \ u_{\bp-})$, where $u_{\bp s}$ is the energy eigenvector, $\bsigma\cdot\bp \,u_{\bp s} = \epsilon_{{\bp}s}u_{\bp s}$ with $\epsilon_{{\bp}s} = s |\bp|$ for $s = \pm$, the amplitude is written as~\cite{Stephanov:2012ki}
\begin{equation}
\begin{split}
\mathcal{A}
&= V_{\bp_\mathrm{f}}\int \mathcal{D} \bx \mathcal{D} \bp\, \mathcal{P} \exp\biggl[\frac{\rmi}{\hbar} \int_{t_\mathrm{i}}^{t_\mathrm{f}} \rmd t (\bp\cdot\dot{\bx} - H)\biggr]V_{\bp_\mathrm{i}}^\dag,
\end{split}
\label{eq:pathint1}
\end{equation}
where $\mathcal{P}$ denotes the path-ordered product of the $\mathrm{U}(2)$ matrices.
Here, the effective Hamiltonian $H$ is expressed as
$H=H_0+\bar{H}_1+\delta H_1$ with $H_0 := |\bp|\sigma_3 +\Phi$, $\bar{H}_1 := \hbar\bar{\ba}\cdot\dot\bp $, and $\delta H_1 := \hbar\delta\ba\cdot\dot{\bp}$,
where $\bar{\ba}=\sigma_3\ba$ and $\delta\ba$ are diagonal and off-diagonal matrix elements of the Berry connection
$\hat{\ba} = -\rmi V_\bp^\dag \bnabla_\bp V_\bp$, respectively.

It should be noted that the integrand is still not diagonalized because of the presence of $\delta H_1$.
If this off-diagonal matrix element is of order ${\cal O}(\hbar^2)$, we obtain the effective Lagrangian describing the conventional chiral kinetic theory up to ${\cal O}(\hbar)$.
The original work in Ref.~\cite{Stephanov:2012ki} motivates the power counting $\delta H_1 ={\cal O}(\hbar^2)$ via the 't Hooft abelian projection~\cite{tHooft:1981bkw}.
In this Letter, employing a more systematic scheme developed by Luttinger-Kohn~\cite{Luttinger-Kohn} and Schrieffer-Wolff~\cite{Schrieffer-Wolff}, we show that $\delta H_1$ can be substituted by an $\mathcal{O}(\hbar^2)$ diagonal matrix, and eventually derive the Lagrangian~\eqref{eq:L} from Eq.~\eqref{eq:pathint1}.

We first introduce a unitary operator $\rme^{S_{\bp}}$, where an anti-Hermitian matrix $S_{\bp}={\cal O}(\hbar)$ fulfills $[S,H_0] + \delta H_1 = 0$.
Let us here compute the matrix element in the chirality basis, $S_{s,s'} = \langle s|S|s'\rangle$.
While the diagonal elements $S_{s,s}$ can be set to zero, the above condition requires
\begin{equation}
\begin{split}
\label{eq:S_off-diag}
 S_{s,-s} = \frac{[\delta H_1]_{s,-s}}{\epsilon_{{\bp}s}-\epsilon_
{{\bp}-s}}
 = s\hbar \frac{\dot{\bp}\cdot \hat{\ba}_{s,-s}}{2|\bp|}.
\end{split}
\end{equation}

In order to perform the perturbative diagonalization of the integrand, we discretize the integrand in Eq.~\eqref{eq:pathint1} with an interval $\Delta t$, leading to an infinite product of exponential factors.
Inserting the identity $\rme^{S}\rme^{-S} = 1$ between each two neighboring exponential factors, we rewrite the amplitude as the product of $\rme^{S_{\bp}} \rme^{-S_{\bp'}}\simeq \rme^{\Delta\bp\cdot\bnabla_\bp S_\bp}$ and
\begin{equation}
\begin{split}
\label{eq:Campbell_S}
  &\rme^{S} \rme^{-\frac{\rmi}{\hbar} H\Delta t} \rme^{-S} 
  \simeq \rme^{
 	-\frac{\rmi}{\hbar}(H_0+\bar{H}_1+[S,\bar{H}_1] + \frac{1}{2}[S,\delta H_1])\Delta t}, 
\end{split}
\end{equation}
where we ignored the ${\cal O}(\hbar^3)$ contribution in the parentheses and defined $\bp'=\bp-\Delta\bp$.
Here, we dropped the ${\cal O}(S^2)$ terms in the exponential (e.g., $-[S_\bp,\bnabla_\bp S_\bp]\cdot\Delta\bp/2$), as they contribute to the effective Lagrangian at ${\cal O}(\hbar^3)$.

Combining all of the exponential factors, we arrive at the following amplitude:
\begin{equation}
\begin{split}
\mathcal{A}
&= U_{\bp_\mathrm{f}}\int \mathcal{D} \bx \mathcal{D} \bp\, \mathcal{P}\exp\biggl[\frac{\rmi}{\hbar}\int_{t_\mathrm{i}}^{t_\mathrm{f}} \hat{L}\,\rmd t \biggr] U^\dag_{\bp_\mathrm{i}},
\end{split}
\end{equation}
where $U=V\rme^{-S}$ and $\hat L = \bp\cdot\dot{\bx} - H_0 - \bar{H}_1 - \frac{\hbar^2}{2}\dot{p}_i\dot{p}_j \hat{G}_{ij}$ with
\begin{equation}
 \hat{G}_{ij}
 := 
  \begin{pmatrix}
  \ds\frac{[\hat{a}_{(i}]_{+-}[\hat{a}_{j)}]_{-+}}{2|\bp|}
  &\ds (\rmi \partial^p_{(i} - 2a_{(i}) \frac{[\hat{a}_{j)}]_{+-}}{2|\bp|} \\[0.5em]
  \ds(-\rmi\partial^p_{(i} - 2a_{(i}) \frac{[\hat{a}_{j)}]_{-+}}{2|\bp|}
  &\ds  -\frac{[\hat{a}_{(i}]_{+-}[\hat{a}_{j)}]_{-+}}{2|\bp|}
 \end{pmatrix}
\end{equation}
and $X_{(i}Y_{j)} :=X_iY_j+X_jY_i$.
Although this Lagrangian still involves the off-diagonal components in $\hat{G}_{ij}$, they can be eliminated up to ${\cal O}(\hbar^2)$ in the same manner as before.
That is, we introduce another unitary matrix $\rme^{T_{\bp}}$ with $T_{\bp}={\cal O}(\hbar^2)$ such that $[T,H_0] + \delta H_2 = 0$, where $\delta H_2$ is the off-diagonal part of $\hat{G}_{ij}$.
We can, however, show that the additional terms arising from this procedure are of order ${\cal O}(\hbar^3)$.
Eventually, the effective Lagrangian corresponding to the ${\cal O}(\hbar^2)$ nonlinear chiral kinetic theory for the positive-helicity state reduces to Eq.~(\ref{eq:L}), where $\epsilon=\epsilon_\bp+\Phi(\bx)$ with $\epsilon_\bp = |\bp|$, and $G_{ij}$ is the diagonal components of $\hat{G}_{ij}$, given by Eqs.~(\ref{eq:Gij}).
We note that, in a similar manner, the $\mathcal{O}(\hbar^3)$ contributions can be systematically identified, and the application of the Dirac-Bergmann theory to the resulting Lagrangian would yield additional geometric structures beyond the Berry curvature and quantum metric~\cite{Kozii,Avdoshkin,Avdoshkin:2024rch,Ahn_2021,Avdoshkin_prl,Mitscherling:2024uxz,Hetenyi:2023hye}.

\sect{Outlook}%
Our work has important implications not only for nonlinear responses in condensed matter systems with nontrivial quantum metric, but also for those in high-energy and astrophysical systems involving Weyl and Dirac fermions. 
In particular, our finding of a quantum-metric correction to the energy density indicates that the equations of state of relativistic many-body systems composed of Dirac fermions, such as quark–gluon plasmas created in heavy-ion collisions or quark matter inside neutron stars, are affected by the quantum metric coupled to electric fields, even in the absence of chirality imbalance. While this study has focused on a static inhomogeneous electric field for simplicity, incorporating time-dependent electric and magnetic fields will be essential for further phenomenological applications.
One may also include dynamical electromagnetic fields in addition to background fields, enabling the study of collective modes with quantum metric in the kinetic regime, as in Ref.~\cite{Akamatsu:2013pjd}.

In this Letter, we focused on particle number and energy densities as well as dissipationless geometric electric and energy currents.
A natural direction would be to study dissipative currents by incorporating collision terms into the kinetic equation. 
From a more theoretical viewpoint, it would be interesting to explore the consequences of Lorentz invariance on the nonlinear chiral kinetic theory with the quantum metric, in the spirit of Refs.~\cite{Son:2012zy,Chen:2014cla}, and 
to develop Lorentz-covariant extensions \cite{Chen:2015gta} of such a kinetic theory; see also Refs.~\cite{Mameda:2023ueq,Hayata2020} for related discussions.
Another direction is the extension to curved spacetime~\cite{Liu:2018xip,Hayata2020},
and the study of its interplay with the quantum metric, which is relevant in cosmology and astrophysics.

\sect{Acknowledgments}%
The authors thank Tomoki Ozawa for discussions.
This work was supported by JSPS KAKENHI Grant Numbers 24K17052 (K.\,M.) and JP24K00631 (N.\,Y.), and by the Ishii-Ishibashi Fund (Keio University Grant for Early Career Researchers).

\bibliographystyle{apsrev4-2}
\bibliography{ckt}

%apsrev4-2.bst 2019-01-14 (MD) hand-edited version of apsrev4-1.bst
%Control: key (0)
%Control: author (72) initials jnrlst
%Control: editor formatted (1) identically to author
%Control: production of article title (-1) disabled
%Control: page (0) single
%Control: year (1) truncated
%Control: production of eprint (0) enabled
\begin{thebibliography}{60}%
\makeatletter
\providecommand \@ifxundefined [1]{%
 \@ifx{#1\undefined}
}%
\providecommand \@ifnum [1]{%
 \ifnum #1\expandafter \@firstoftwo
 \else \expandafter \@secondoftwo
 \fi
}%
\providecommand \@ifx [1]{%
 \ifx #1\expandafter \@firstoftwo
 \else \expandafter \@secondoftwo
 \fi
}%
\providecommand \natexlab [1]{#1}%
\providecommand \enquote  [1]{``#1''}%
\providecommand \bibnamefont  [1]{#1}%
\providecommand \bibfnamefont [1]{#1}%
\providecommand \citenamefont [1]{#1}%
\providecommand \href@noop [0]{\@secondoftwo}%
\providecommand \href [0]{\begingroup \@sanitize@url \@href}%
\providecommand \@href[1]{\@@startlink{#1}\@@href}%
\providecommand \@@href[1]{\endgroup#1\@@endlink}%
\providecommand \@sanitize@url [0]{\catcode `\\12\catcode `\$12\catcode
  `\&12\catcode `\#12\catcode `\^12\catcode `\_12\catcode `\%12\relax}%
\providecommand \@@startlink[1]{}%
\providecommand \@@endlink[0]{}%
\providecommand \url  [0]{\begingroup\@sanitize@url \@url }%
\providecommand \@url [1]{\endgroup\@href {#1}{\urlprefix }}%
\providecommand \urlprefix  [0]{URL }%
\providecommand \Eprint [0]{\href }%
\providecommand \doibase [0]{https://doi.org/}%
\providecommand \selectlanguage [0]{\@gobble}%
\providecommand \bibinfo  [0]{\@secondoftwo}%
\providecommand \bibfield  [0]{\@secondoftwo}%
\providecommand \translation [1]{[#1]}%
\providecommand \BibitemOpen [0]{}%
\providecommand \bibitemStop [0]{}%
\providecommand \bibitemNoStop [0]{.\EOS\space}%
\providecommand \EOS [0]{\spacefactor3000\relax}%
\providecommand \BibitemShut  [1]{\csname bibitem#1\endcsname}%
\let\auto@bib@innerbib\@empty
%</preamble>
\bibitem [{\citenamefont {Berry}(1984)}]{Berry}%
  \BibitemOpen
  \bibfield  {author} {\bibinfo {author} {\bibfnamefont {M.~V.}\ \bibnamefont
  {Berry}},\ }\href {https://doi.org/10.1098/rspa.1984.0023} {\bibfield
  {journal} {\bibinfo  {journal} {Proc. R. Soc. London A}\ }\textbf {\bibinfo
  {volume} {392}},\ \bibinfo {pages} {45} (\bibinfo {year} {1984})}\BibitemShut
  {NoStop}%
\bibitem [{\citenamefont {Xiao}\ \emph {et~al.}(2010)\citenamefont {Xiao},
  \citenamefont {Chang},\ and\ \citenamefont {Niu}}]{Xiao2010}%
  \BibitemOpen
  \bibfield  {author} {\bibinfo {author} {\bibfnamefont {D.}~\bibnamefont
  {Xiao}}, \bibinfo {author} {\bibfnamefont {M.-C.}\ \bibnamefont {Chang}},\
  and\ \bibinfo {author} {\bibfnamefont {Q.}~\bibnamefont {Niu}},\ }\href
  {https://doi.org/10.1103/RevModPhys.82.1959} {\bibfield  {journal} {\bibinfo
  {journal} {Rev. Mod. Phys.}\ }\textbf {\bibinfo {volume} {82}},\ \bibinfo
  {pages} {1959} (\bibinfo {year} {2010})},\ \Eprint
  {https://arxiv.org/abs/0907.2021} {arXiv:0907.2021 [cond-mat.mes-hall]}
  \BibitemShut {NoStop}%
\bibitem [{\citenamefont {Kamada}\ \emph {et~al.}(2023)\citenamefont {Kamada},
  \citenamefont {Yamamoto},\ and\ \citenamefont {Yang}}]{Kamada:2022nyt}%
  \BibitemOpen
  \bibfield  {author} {\bibinfo {author} {\bibfnamefont {K.}~\bibnamefont
  {Kamada}}, \bibinfo {author} {\bibfnamefont {N.}~\bibnamefont {Yamamoto}},\
  and\ \bibinfo {author} {\bibfnamefont {D.-L.}\ \bibnamefont {Yang}},\ }\href
  {https://doi.org/10.1016/j.ppnp.2022.104016} {\bibfield  {journal} {\bibinfo
  {journal} {Prog. Part. Nucl. Phys.}\ }\textbf {\bibinfo {volume} {129}},\
  \bibinfo {pages} {104016} (\bibinfo {year} {2023})},\ \Eprint
  {https://arxiv.org/abs/2207.09184} {arXiv:2207.09184 [astro-ph.CO]}
  \BibitemShut {NoStop}%
\bibitem [{\citenamefont {Son}\ and\ \citenamefont
  {Yamamoto}(2012)}]{Son:2012wh}%
  \BibitemOpen
  \bibfield  {author} {\bibinfo {author} {\bibfnamefont {D.~T.}\ \bibnamefont
  {Son}}\ and\ \bibinfo {author} {\bibfnamefont {N.}~\bibnamefont {Yamamoto}},\
  }\href {https://doi.org/10.1103/PhysRevLett.109.181602} {\bibfield  {journal}
  {\bibinfo  {journal} {Phys. Rev. Lett.}\ }\textbf {\bibinfo {volume} {109}},\
  \bibinfo {pages} {181602} (\bibinfo {year} {2012})},\ \Eprint
  {https://arxiv.org/abs/1203.2697} {arXiv:1203.2697 [cond-mat.mes-hall]}
  \BibitemShut {NoStop}%
\bibitem [{\citenamefont {Stephanov}\ and\ \citenamefont
  {Yin}(2012)}]{Stephanov:2012ki}%
  \BibitemOpen
  \bibfield  {author} {\bibinfo {author} {\bibfnamefont {M.~A.}\ \bibnamefont
  {Stephanov}}\ and\ \bibinfo {author} {\bibfnamefont {Y.}~\bibnamefont
  {Yin}},\ }\href {https://doi.org/10.1103/PhysRevLett.109.162001} {\bibfield
  {journal} {\bibinfo  {journal} {Phys. Rev. Lett.}\ }\textbf {\bibinfo
  {volume} {109}},\ \bibinfo {pages} {162001} (\bibinfo {year} {2012})},\
  \Eprint {https://arxiv.org/abs/1207.0747} {arXiv:1207.0747 [hep-th]}
  \BibitemShut {NoStop}%
\bibitem [{\citenamefont {Son}\ and\ \citenamefont
  {Yamamoto}(2013)}]{Son:2012zy}%
  \BibitemOpen
  \bibfield  {author} {\bibinfo {author} {\bibfnamefont {D.~T.}\ \bibnamefont
  {Son}}\ and\ \bibinfo {author} {\bibfnamefont {N.}~\bibnamefont {Yamamoto}},\
  }\href {https://doi.org/10.1103/PhysRevD.87.085016} {\bibfield  {journal}
  {\bibinfo  {journal} {Phys. Rev. D}\ }\textbf {\bibinfo {volume} {87}},\
  \bibinfo {pages} {085016} (\bibinfo {year} {2013})},\ \Eprint
  {https://arxiv.org/abs/1210.8158} {arXiv:1210.8158 [hep-th]} \BibitemShut
  {NoStop}%
\bibitem [{\citenamefont {Vilenkin}(1980)}]{Vilenkin:1980fu}%
  \BibitemOpen
  \bibfield  {author} {\bibinfo {author} {\bibfnamefont {A.}~\bibnamefont
  {Vilenkin}},\ }\href {https://doi.org/10.1103/PhysRevD.22.3080} {\bibfield
  {journal} {\bibinfo  {journal} {Phys. Rev.}\ }\textbf {\bibinfo {volume}
  {D22}},\ \bibinfo {pages} {3080} (\bibinfo {year} {1980})}\BibitemShut
  {NoStop}%
\bibitem [{\citenamefont {Nielsen}\ and\ \citenamefont
  {Ninomiya}(1983)}]{Nielsen:1983rb}%
  \BibitemOpen
  \bibfield  {author} {\bibinfo {author} {\bibfnamefont {H.~B.}\ \bibnamefont
  {Nielsen}}\ and\ \bibinfo {author} {\bibfnamefont {M.}~\bibnamefont
  {Ninomiya}},\ }\href {https://doi.org/10.1016/0370-2693(83)91529-0}
  {\bibfield  {journal} {\bibinfo  {journal} {Phys. Lett.}\ }\textbf {\bibinfo
  {volume} {B130}},\ \bibinfo {pages} {389} (\bibinfo {year}
  {1983})}\BibitemShut {NoStop}%
\bibitem [{\citenamefont {Fukushima}\ \emph {et~al.}(2008)\citenamefont
  {Fukushima}, \citenamefont {Kharzeev},\ and\ \citenamefont
  {Warringa}}]{Fukushima:2008xe}%
  \BibitemOpen
  \bibfield  {author} {\bibinfo {author} {\bibfnamefont {K.}~\bibnamefont
  {Fukushima}}, \bibinfo {author} {\bibfnamefont {D.~E.}\ \bibnamefont
  {Kharzeev}},\ and\ \bibinfo {author} {\bibfnamefont {H.~J.}\ \bibnamefont
  {Warringa}},\ }\href {https://doi.org/10.1103/PhysRevD.78.074033} {\bibfield
  {journal} {\bibinfo  {journal} {Phys. Rev.}\ }\textbf {\bibinfo {volume}
  {D78}},\ \bibinfo {pages} {074033} (\bibinfo {year} {2008})},\ \Eprint
  {https://arxiv.org/abs/0808.3382} {arXiv:0808.3382 [hep-ph]} \BibitemShut
  {NoStop}%
\bibitem [{\citenamefont {Adler}(1969)}]{Adler:1969gk}%
  \BibitemOpen
  \bibfield  {author} {\bibinfo {author} {\bibfnamefont {S.~L.}\ \bibnamefont
  {Adler}},\ }\href {https://doi.org/10.1103/PhysRev.177.2426} {\bibfield
  {journal} {\bibinfo  {journal} {Phys. Rev.}\ }\textbf {\bibinfo {volume}
  {177}},\ \bibinfo {pages} {2426} (\bibinfo {year} {1969})}\BibitemShut
  {NoStop}%
\bibitem [{\citenamefont {Bell}\ and\ \citenamefont
  {Jackiw}(1969)}]{Bell:1969ts}%
  \BibitemOpen
  \bibfield  {author} {\bibinfo {author} {\bibfnamefont {J.~S.}\ \bibnamefont
  {Bell}}\ and\ \bibinfo {author} {\bibfnamefont {R.}~\bibnamefont {Jackiw}},\
  }\href {https://doi.org/10.1007/BF02823296} {\bibfield  {journal} {\bibinfo
  {journal} {Nuovo Cimento A}\ }\textbf {\bibinfo {volume} {60}},\ \bibinfo
  {pages} {47} (\bibinfo {year} {1969})}\BibitemShut {NoStop}%
\bibitem [{\citenamefont {Provost}\ and\ \citenamefont
  {Vall\'ee}(1980)}]{Provost1980}%
  \BibitemOpen
  \bibfield  {author} {\bibinfo {author} {\bibfnamefont {J.~P.}\ \bibnamefont
  {Provost}}\ and\ \bibinfo {author} {\bibfnamefont {G.}~\bibnamefont
  {Vall\'ee}},\ }\href {https://doi.org/10.1007/BF02193559} {\bibfield
  {journal} {\bibinfo  {journal} {Commun. Math. Phys.}\ }\textbf {\bibinfo
  {volume} {76}},\ \bibinfo {pages} {289} (\bibinfo {year} {1980})}\BibitemShut
  {NoStop}%
\bibitem [{\citenamefont {Liu}\ \emph {et~al.}(2024)\citenamefont {Liu},
  \citenamefont {Qiang}, \citenamefont {Lu},\ and\ \citenamefont
  {Xie}}]{Liu_2024}%
  \BibitemOpen
  \bibfield  {author} {\bibinfo {author} {\bibfnamefont {T.}~\bibnamefont
  {Liu}}, \bibinfo {author} {\bibfnamefont {X.-B.}\ \bibnamefont {Qiang}},
  \bibinfo {author} {\bibfnamefont {H.-Z.}\ \bibnamefont {Lu}},\ and\ \bibinfo
  {author} {\bibfnamefont {X.~C.}\ \bibnamefont {Xie}},\ }\href
  {https://doi.org/10.1093/nsr/nwae334} {\bibfield  {journal} {\bibinfo
  {journal} {Natl. Sci. Rev.}\ }\textbf {\bibinfo {volume} {12}},\ \bibinfo
  {pages} {nwae334} (\bibinfo {year} {2024})},\ \Eprint
  {https://arxiv.org/abs/2409.13408} {arXiv:2409.13408 [cond-mat.mes-hall]}
  \BibitemShut {NoStop}%
\bibitem [{\citenamefont {Gao}\ \emph {et~al.}(2025)\citenamefont {Gao},
  \citenamefont {Nagaosa}, \citenamefont {Ni},\ and\ \citenamefont
  {Xu}}]{Gao2025}%
  \BibitemOpen
  \bibfield  {author} {\bibinfo {author} {\bibfnamefont {A.}~\bibnamefont
  {Gao}}, \bibinfo {author} {\bibfnamefont {N.}~\bibnamefont {Nagaosa}},
  \bibinfo {author} {\bibfnamefont {N.}~\bibnamefont {Ni}},\ and\ \bibinfo
  {author} {\bibfnamefont {S.-Y.}\ \bibnamefont {Xu}},\ }\href@noop {}
  {\bibinfo {title} {Quantum geometry phenomena in condensed matter systems}}
  (\bibinfo {year} {2025}),\ \Eprint {https://arxiv.org/abs/2508.00469}
  {arXiv:2508.00469 [cond-mat.str-el]} \BibitemShut {NoStop}%
\bibitem [{\citenamefont {Verma}\ \emph {et~al.}(2025)\citenamefont {Verma},
  \citenamefont {Moll}, \citenamefont {Holder},\ and\ \citenamefont
  {Queiroz}}]{Verma}%
  \BibitemOpen
  \bibfield  {author} {\bibinfo {author} {\bibfnamefont {N.}~\bibnamefont
  {Verma}}, \bibinfo {author} {\bibfnamefont {P.~J.~W.}\ \bibnamefont {Moll}},
  \bibinfo {author} {\bibfnamefont {T.}~\bibnamefont {Holder}},\ and\ \bibinfo
  {author} {\bibfnamefont {R.}~\bibnamefont {Queiroz}},\ }\href
  {https://arxiv.org/abs/2504.07173} {\bibinfo {title} {Quantum geometry:
  Revisiting electronic scales in quantum matter}} (\bibinfo {year} {2025}),\
  \Eprint {https://arxiv.org/abs/2504.07173} {arXiv:2504.07173
  [cond-mat.mtrl-sci]} \BibitemShut {NoStop}%
\bibitem [{\citenamefont {Jiang}\ \emph {et~al.}(2025)\citenamefont {Jiang},
  \citenamefont {Holder},\ and\ \citenamefont {Yan}}]{Jiang_2025}%
  \BibitemOpen
  \bibfield  {author} {\bibinfo {author} {\bibfnamefont {Y.}~\bibnamefont
  {Jiang}}, \bibinfo {author} {\bibfnamefont {T.}~\bibnamefont {Holder}},\ and\
  \bibinfo {author} {\bibfnamefont {B.}~\bibnamefont {Yan}},\ }\href
  {https://doi.org/10.1088/1361-6633/ade454} {\bibfield  {journal} {\bibinfo
  {journal} {Reports on Progress in Physics}\ }\textbf {\bibinfo {volume}
  {88}},\ \bibinfo {pages} {076502} (\bibinfo {year} {2025})}\BibitemShut
  {NoStop}%
\bibitem [{\citenamefont {Yu}\ \emph {et~al.}(2025)\citenamefont {Yu},
  \citenamefont {Bernevig}, \citenamefont {Queiroz}, \citenamefont {Rossi},
  \citenamefont {T{\"o}rm{\"a}},\ and\ \citenamefont {Yang}}]{yu2025quantum}%
  \BibitemOpen
  \bibfield  {author} {\bibinfo {author} {\bibfnamefont {J.}~\bibnamefont
  {Yu}}, \bibinfo {author} {\bibfnamefont {B.~A.}\ \bibnamefont {Bernevig}},
  \bibinfo {author} {\bibfnamefont {R.}~\bibnamefont {Queiroz}}, \bibinfo
  {author} {\bibfnamefont {E.}~\bibnamefont {Rossi}}, \bibinfo {author}
  {\bibfnamefont {P.}~\bibnamefont {T{\"o}rm{\"a}}},\ and\ \bibinfo {author}
  {\bibfnamefont {B.-J.}\ \bibnamefont {Yang}},\ }\href
  {https://doi.org/10.1038/s41535-025-00801-3} {\bibfield  {journal} {\bibinfo
  {journal} {npj Quantum Materials}\ }\textbf {\bibinfo {volume} {10}},\
  \bibinfo {pages} {101} (\bibinfo {year} {2025})}\BibitemShut {NoStop}%
\bibitem [{\citenamefont {Xiao}\ \emph {et~al.}(2005)\citenamefont {Xiao},
  \citenamefont {Shi},\ and\ \citenamefont {Niu}}]{Xiao:2005qw}%
  \BibitemOpen
  \bibfield  {author} {\bibinfo {author} {\bibfnamefont {D.}~\bibnamefont
  {Xiao}}, \bibinfo {author} {\bibfnamefont {J.}~\bibnamefont {Shi}},\ and\
  \bibinfo {author} {\bibfnamefont {Q.}~\bibnamefont {Niu}},\ }\href
  {https://doi.org/10.1103/PhysRevLett.95.137204} {\bibfield  {journal}
  {\bibinfo  {journal} {Phys. Rev. Lett.}\ }\textbf {\bibinfo {volume} {95}},\
  \bibinfo {pages} {137204} (\bibinfo {year} {2005})},\ \Eprint
  {https://arxiv.org/abs/cond-mat/0502340} {cond-mat/0502340} \BibitemShut
  {NoStop}%
\bibitem [{\citenamefont {Duval}\ \emph {et~al.}(2006)\citenamefont {Duval},
  \citenamefont {Horvath}, \citenamefont {Horvathy}, \citenamefont {Martina},\
  and\ \citenamefont {Stichel}}]{Duval:2005vn}%
  \BibitemOpen
  \bibfield  {author} {\bibinfo {author} {\bibfnamefont {C.}~\bibnamefont
  {Duval}}, \bibinfo {author} {\bibfnamefont {Z.}~\bibnamefont {Horvath}},
  \bibinfo {author} {\bibfnamefont {P.~A.}\ \bibnamefont {Horvathy}}, \bibinfo
  {author} {\bibfnamefont {L.}~\bibnamefont {Martina}},\ and\ \bibinfo {author}
  {\bibfnamefont {P.}~\bibnamefont {Stichel}},\ }\href
  {https://doi.org/10.1142/S0217984906010573} {\bibfield  {journal} {\bibinfo
  {journal} {Mod. Phys. Lett. B}\ }\textbf {\bibinfo {volume} {20}},\ \bibinfo
  {pages} {373} (\bibinfo {year} {2006})},\ \Eprint
  {https://arxiv.org/abs/cond-mat/0506051} {cond-mat/0506051} \BibitemShut
  {NoStop}%
\bibitem [{\citenamefont {Dirac}(1950)}]{Dirac1950}%
  \BibitemOpen
  \bibfield  {author} {\bibinfo {author} {\bibfnamefont {P.~A.~M.}\
  \bibnamefont {Dirac}},\ }\href {https://doi.org/10.4153/CJM-1950-012-1}
  {\bibfield  {journal} {\bibinfo  {journal} {Can. J. Math}\ }\textbf {\bibinfo
  {volume} {2}},\ \bibinfo {pages} {129} (\bibinfo {year} {1950})}\BibitemShut
  {NoStop}%
\bibitem [{\citenamefont {Dirac}(1964)}]{Dirac}%
  \BibitemOpen
  \bibfield  {author} {\bibinfo {author} {\bibfnamefont {P.~A.~M.}\
  \bibnamefont {Dirac}},\ }\href@noop {} {\emph {\bibinfo {title} {Lectures on
  Quantum Mechanics}}}\ (\bibinfo  {publisher} {Belfer Graduate School of
  Science, Yeshiva University, New York},\ \bibinfo {year} {1964})\BibitemShut
  {NoStop}%
\bibitem [{\citenamefont {Gao}\ \emph {et~al.}(2014)\citenamefont {Gao},
  \citenamefont {Yang},\ and\ \citenamefont {Niu}}]{Gao:2014}%
  \BibitemOpen
  \bibfield  {author} {\bibinfo {author} {\bibfnamefont {Y.}~\bibnamefont
  {Gao}}, \bibinfo {author} {\bibfnamefont {S.~A.}\ \bibnamefont {Yang}},\ and\
  \bibinfo {author} {\bibfnamefont {Q.}~\bibnamefont {Niu}},\ }\href
  {https://doi.org/10.1103/PhysRevLett.112.166601} {\bibfield  {journal}
  {\bibinfo  {journal} {Phys. Rev. Lett.}\ }\textbf {\bibinfo {volume} {112}},\
  \bibinfo {pages} {166601} (\bibinfo {year} {2014})},\ \Eprint
  {https://arxiv.org/abs/1402.2538} {arXiv:1402.2538 [cond-mat.mes-hall]}
  \BibitemShut {NoStop}%
\bibitem [{\citenamefont {Kaplan}\ \emph {et~al.}(2024)\citenamefont {Kaplan},
  \citenamefont {Holder},\ and\ \citenamefont {Yan}}]{Kaplan:2024}%
  \BibitemOpen
  \bibfield  {author} {\bibinfo {author} {\bibfnamefont {D.}~\bibnamefont
  {Kaplan}}, \bibinfo {author} {\bibfnamefont {T.}~\bibnamefont {Holder}},\
  and\ \bibinfo {author} {\bibfnamefont {B.}~\bibnamefont {Yan}},\ }\href
  {https://doi.org/10.1103/PhysRevLett.132.026301} {\bibfield  {journal}
  {\bibinfo  {journal} {Phys. Rev. Lett.}\ }\textbf {\bibinfo {volume} {132}},\
  \bibinfo {pages} {026301} (\bibinfo {year} {2024})},\ \Eprint
  {https://arxiv.org/abs/2211.17213} {arXiv:2211.17213 [cond-mat.mes-hall]}
  \BibitemShut {NoStop}%
\bibitem [{\citenamefont {Das}\ \emph {et~al.}(2023)\citenamefont {Das},
  \citenamefont {Lahiri}, \citenamefont {Atencia}, \citenamefont {Culcer},\
  and\ \citenamefont {Agarwal}}]{Das}%
  \BibitemOpen
  \bibfield  {author} {\bibinfo {author} {\bibfnamefont {K.}~\bibnamefont
  {Das}}, \bibinfo {author} {\bibfnamefont {S.}~\bibnamefont {Lahiri}},
  \bibinfo {author} {\bibfnamefont {R.~B.}\ \bibnamefont {Atencia}}, \bibinfo
  {author} {\bibfnamefont {D.}~\bibnamefont {Culcer}},\ and\ \bibinfo {author}
  {\bibfnamefont {A.}~\bibnamefont {Agarwal}},\ }\href
  {https://doi.org/10.1103/PhysRevB.108.L201405} {\bibfield  {journal}
  {\bibinfo  {journal} {Phys. Rev. B}\ }\textbf {\bibinfo {volume} {108}},\
  \bibinfo {pages} {L201405} (\bibinfo {year} {2023})}\BibitemShut {NoStop}%
\bibitem [{\citenamefont {Hidaka}\ \emph {et~al.}(2022)\citenamefont {Hidaka},
  \citenamefont {Pu}, \citenamefont {Wang},\ and\ \citenamefont
  {Yang}}]{Hidaka:2022dmn}%
  \BibitemOpen
  \bibfield  {author} {\bibinfo {author} {\bibfnamefont {Y.}~\bibnamefont
  {Hidaka}}, \bibinfo {author} {\bibfnamefont {S.}~\bibnamefont {Pu}}, \bibinfo
  {author} {\bibfnamefont {Q.}~\bibnamefont {Wang}},\ and\ \bibinfo {author}
  {\bibfnamefont {D.-L.}\ \bibnamefont {Yang}},\ }\href
  {https://doi.org/10.1016/j.ppnp.2022.103989} {\bibfield  {journal} {\bibinfo
  {journal} {Prog. Part. Nucl. Phys.}\ }\textbf {\bibinfo {volume} {127}},\
  \bibinfo {pages} {103989} (\bibinfo {year} {2022})},\ \Eprint
  {https://arxiv.org/abs/2201.07644} {arXiv:2201.07644 [hep-ph]} \BibitemShut
  {NoStop}%
\bibitem [{\citenamefont {Kolodrubetz}\ \emph {et~al.}(2017)\citenamefont
  {Kolodrubetz}, \citenamefont {Sels}, \citenamefont {Mehta},\ and\
  \citenamefont {Polkovnikov}}]{Kolodrubetz2017}%
  \BibitemOpen
  \bibfield  {author} {\bibinfo {author} {\bibfnamefont {M.}~\bibnamefont
  {Kolodrubetz}}, \bibinfo {author} {\bibfnamefont {D.}~\bibnamefont {Sels}},
  \bibinfo {author} {\bibfnamefont {P.}~\bibnamefont {Mehta}},\ and\ \bibinfo
  {author} {\bibfnamefont {A.}~\bibnamefont {Polkovnikov}},\ }\href
  {https://doi.org/10.1016/j.physrep.2017.07.001} {\bibfield  {journal}
  {\bibinfo  {journal} {Phys. Rep.}\ }\textbf {\bibinfo {volume} {697}},\
  \bibinfo {pages} {1} (\bibinfo {year} {2017})},\ \Eprint
  {https://arxiv.org/abs/1602.01062} {arXiv:1602.01062 [cond-mat.quant-gas]}
  \BibitemShut {NoStop}%
\bibitem [{\citenamefont {Ren}(2025)}]{Ren:2025zei}%
  \BibitemOpen
  \bibfield  {author} {\bibinfo {author} {\bibfnamefont {Y.}~\bibnamefont
  {Ren}},\ }\href@noop {} {\bibinfo {title} {{Momentum-space Metric Tensor from
  Nonadiabatic Evolution of Bloch Electrons}}} (\bibinfo {year} {2025}),\
  \Eprint {https://arxiv.org/abs/2506.06439} {arXiv:2506.06439
  [cond-mat.mtrl-sci]} \BibitemShut {NoStop}%
\bibitem [{\citenamefont {Yoshida}\ and\ \citenamefont
  {Yokoyama}(2025)}]{Yoshida:2025tuo}%
  \BibitemOpen
  \bibfield  {author} {\bibinfo {author} {\bibfnamefont {H.}~\bibnamefont
  {Yoshida}}\ and\ \bibinfo {author} {\bibfnamefont {T.}~\bibnamefont
  {Yokoyama}},\ }\href@noop {} {\bibinfo {title} {{Emergent-gravity Hall effect
  from quantum geometry}}} (\bibinfo {year} {2025}),\ \Eprint
  {https://arxiv.org/abs/2507.18458} {arXiv:2507.18458 [cond-mat.other]}
  \BibitemShut {NoStop}%
\bibitem [{Note1()}]{Note1}%
  \BibitemOpen
  \bibinfo {note} {This is an effective Lagrangian for a specific band, with
  interband contributions encoded in $G_{ij}$ and ${\protect \bm {a}}$; see the
  derivation for chiral fermions below.}\BibitemShut {Stop}%
\bibitem [{\citenamefont {Jain}\ \emph {et~al.}(2025)\citenamefont {Jain},
  \citenamefont {Jankowski},\ and\ \citenamefont {Slager}}]{Jain2025}%
  \BibitemOpen
  \bibfield  {author} {\bibinfo {author} {\bibfnamefont {A.}~\bibnamefont
  {Jain}}, \bibinfo {author} {\bibfnamefont {W.~J.}\ \bibnamefont
  {Jankowski}},\ and\ \bibinfo {author} {\bibfnamefont {R.-J.}\ \bibnamefont
  {Slager}},\ }\href {https://doi.org/10.1103/5gmg-q1z5} {\bibfield  {journal}
  {\bibinfo  {journal} {Phys. Rev. B}\ }\textbf {\bibinfo {volume} {111}},\
  \bibinfo {pages} {235149} (\bibinfo {year} {2025})}\BibitemShut {NoStop}%
\bibitem [{\citenamefont {Fontana}\ \emph {et~al.}(2025)\citenamefont
  {Fontana}, \citenamefont {Velasco}, \citenamefont {Niu}, \citenamefont {Ye},
  \citenamefont {Lopes}, \citenamefont {de~Souza}, \citenamefont {Moutinho},
  \citenamefont {Lewenkopf},\ and\ \citenamefont {Neto}}]{Fontana2025}%
  \BibitemOpen
  \bibfield  {author} {\bibinfo {author} {\bibfnamefont {P.}~\bibnamefont
  {Fontana}}, \bibinfo {author} {\bibfnamefont {V.}~\bibnamefont {Velasco}},
  \bibinfo {author} {\bibfnamefont {C.}~\bibnamefont {Niu}}, \bibinfo {author}
  {\bibfnamefont {P.~D.}\ \bibnamefont {Ye}}, \bibinfo {author} {\bibfnamefont
  {P.~V.}\ \bibnamefont {Lopes}}, \bibinfo {author} {\bibfnamefont {K.~E.~M.}\
  \bibnamefont {de~Souza}}, \bibinfo {author} {\bibfnamefont {M.~V.~O.}\
  \bibnamefont {Moutinho}}, \bibinfo {author} {\bibfnamefont {C.}~\bibnamefont
  {Lewenkopf}},\ and\ \bibinfo {author} {\bibfnamefont {M.~B.~S.}\ \bibnamefont
  {Neto}},\ }\href {https://doi.org/10.1103/7nxc-j62y} {\bibfield  {journal}
  {\bibinfo  {journal} {Phys. Rev. Lett.}\ }\textbf {\bibinfo {volume} {135}},\
  \bibinfo {pages} {106602} (\bibinfo {year} {2025})}\BibitemShut {NoStop}%
\bibitem [{\citenamefont {Mehraeen}(2024)}]{Mehraeen_2024}%
  \BibitemOpen
  \bibfield  {author} {\bibinfo {author} {\bibfnamefont {M.}~\bibnamefont
  {Mehraeen}},\ }\href {https://link.aps.org/doi/10.1103/PhysRevB.110.174423}
  {\bibfield  {journal} {\bibinfo  {journal} {Phys. Rev. B}\ }\textbf {\bibinfo
  {volume} {110}} (\bibinfo {year} {2024})}\BibitemShut {NoStop}%
\bibitem [{\citenamefont {Bergmann}(1949)}]{Bergmann1949}%
  \BibitemOpen
  \bibfield  {author} {\bibinfo {author} {\bibfnamefont {P.~G.}\ \bibnamefont
  {Bergmann}},\ }\href {https://doi.org/10.1103/PhysRev.75.680} {\bibfield
  {journal} {\bibinfo  {journal} {Phys. Rev.}\ }\textbf {\bibinfo {volume}
  {75}},\ \bibinfo {pages} {680} (\bibinfo {year} {1949})}\BibitemShut
  {NoStop}%
\bibitem [{\citenamefont {Faddeev}(1969)}]{Faddeev:1969su}%
  \BibitemOpen
  \bibfield  {author} {\bibinfo {author} {\bibfnamefont {L.~D.}\ \bibnamefont
  {Faddeev}},\ }\href {https://doi.org/10.1007/BF01028566} {\bibfield
  {journal} {\bibinfo  {journal} {Theor. Math. Phys.}\ }\textbf {\bibinfo
  {volume} {1}},\ \bibinfo {pages} {1} (\bibinfo {year} {1969})}\BibitemShut
  {NoStop}%
\bibitem [{Note2()}]{Note2}%
  \BibitemOpen
  \bibinfo {note} {The usefulness of the Dirac bracket has already been
  recognized in the seminal work of the kinetic theory with the Berry
  curvature, Ref.~\cite {Sundaram-Niu}, though not explicitly
  investigated.}\BibitemShut {Stop}%
\bibitem [{Note3()}]{Note3}%
  \BibitemOpen
  \bibinfo {note} {In the standard construction, the time evolution is
  described not by the original Hamiltonian $\protect \tilde {H}$ but by the
  total Hamiltonian $\protect \tilde {H}_\protect \mathrm {T}:=\protect \tilde
  {H}+u_i^x\phi _i^x+u_i^q\phi _i^q$ with Lagrange multipliers $u_i^x$ and
  $u_i^q$. In the present case, however, one can show that the Poisson brackets
  between all the primary constraints vanish, so that the time evolution can be
  described in terms of $\protect \tilde {H}$.}\BibitemShut {Stop}%
\bibitem [{\citenamefont {Liu}\ \emph {et~al.}(2022)\citenamefont {Liu},
  \citenamefont {Zhao}, \citenamefont {Huang}, \citenamefont {Feng},
  \citenamefont {Xiao}, \citenamefont {Wu}, \citenamefont {Lai}, \citenamefont
  {Gao},\ and\ \citenamefont {Yang}}]{Liu2022}%
  \BibitemOpen
  \bibfield  {author} {\bibinfo {author} {\bibfnamefont {H.}~\bibnamefont
  {Liu}}, \bibinfo {author} {\bibfnamefont {J.}~\bibnamefont {Zhao}}, \bibinfo
  {author} {\bibfnamefont {Y.-X.}\ \bibnamefont {Huang}}, \bibinfo {author}
  {\bibfnamefont {X.}~\bibnamefont {Feng}}, \bibinfo {author} {\bibfnamefont
  {C.}~\bibnamefont {Xiao}}, \bibinfo {author} {\bibfnamefont {W.}~\bibnamefont
  {Wu}}, \bibinfo {author} {\bibfnamefont {S.}~\bibnamefont {Lai}}, \bibinfo
  {author} {\bibfnamefont {W.-b.}\ \bibnamefont {Gao}},\ and\ \bibinfo {author}
  {\bibfnamefont {S.~A.}\ \bibnamefont {Yang}},\ }\href
  {https://doi.org/10.1103/PhysRevB.105.045118} {\bibfield  {journal} {\bibinfo
   {journal} {Phys. Rev. B}\ }\textbf {\bibinfo {volume} {105}},\ \bibinfo
  {pages} {045118} (\bibinfo {year} {2022})},\ \Eprint
  {https://arxiv.org/abs/2106.04931} {arXiv:2106.04931 [cond-mat.mes-hall]}
  \BibitemShut {NoStop}%
\bibitem [{\citenamefont {Jia}\ \emph {et~al.}(2024)\citenamefont {Jia},
  \citenamefont {Xiang}, \citenamefont {Qiao},\ and\ \citenamefont
  {Wang}}]{Jia2024}%
  \BibitemOpen
  \bibfield  {author} {\bibinfo {author} {\bibfnamefont {J.}~\bibnamefont
  {Jia}}, \bibinfo {author} {\bibfnamefont {L.}~\bibnamefont {Xiang}}, \bibinfo
  {author} {\bibfnamefont {Z.}~\bibnamefont {Qiao}},\ and\ \bibinfo {author}
  {\bibfnamefont {J.}~\bibnamefont {Wang}},\ }\href
  {https://doi.org/10.1103/PhysRevB.110.245406} {\bibfield  {journal} {\bibinfo
   {journal} {Phys. Rev. B}\ }\textbf {\bibinfo {volume} {110}},\ \bibinfo
  {pages} {245406} (\bibinfo {year} {2024})},\ \Eprint
  {https://arxiv.org/abs/2404.17086} {arXiv:2404.17086 [cond-mat.mes-hall]}
  \BibitemShut {NoStop}%
\bibitem [{\citenamefont {Gao}\ and\ \citenamefont {Xiao}(2019)}]{Gao2019}%
  \BibitemOpen
  \bibfield  {author} {\bibinfo {author} {\bibfnamefont {Y.}~\bibnamefont
  {Gao}}\ and\ \bibinfo {author} {\bibfnamefont {D.}~\bibnamefont {Xiao}},\
  }\href {https://doi.org/10.1103/PhysRevLett.122.227402} {\bibfield  {journal}
  {\bibinfo  {journal} {Phys. Rev. Lett.}\ }\textbf {\bibinfo {volume} {122}},\
  \bibinfo {pages} {227402} (\bibinfo {year} {2019})},\ \Eprint
  {https://arxiv.org/abs/arXiv:1810.02728} {arXiv:arXiv:1810.02728
  [cond-mat.mes-hall]} \BibitemShut {NoStop}%
\bibitem [{\citenamefont {Lapa}\ and\ \citenamefont {Hughes}(2019)}]{Lapa2019}%
  \BibitemOpen
  \bibfield  {author} {\bibinfo {author} {\bibfnamefont {M.~F.}\ \bibnamefont
  {Lapa}}\ and\ \bibinfo {author} {\bibfnamefont {T.~L.}\ \bibnamefont
  {Hughes}},\ }\href {https://doi.org/10.1103/PhysRevB.99.121111} {\bibfield
  {journal} {\bibinfo  {journal} {Phys. Rev. B}\ }\textbf {\bibinfo {volume}
  {99}},\ \bibinfo {pages} {121111} (\bibinfo {year} {2019})},\ \Eprint
  {https://arxiv.org/abs/1812.10497} {arXiv:1812.10497 [cond-mat.mes-hall]}
  \BibitemShut {NoStop}%
\bibitem [{\citenamefont {Qiang}\ \emph {et~al.}(2025)\citenamefont {Qiang},
  \citenamefont {Liu}, \citenamefont {Gao}, \citenamefont {Lu},\ and\
  \citenamefont {Xie}}]{Qiang_2025}%
  \BibitemOpen
  \bibfield  {author} {\bibinfo {author} {\bibfnamefont {X.}~\bibnamefont
  {Qiang}}, \bibinfo {author} {\bibfnamefont {T.}~\bibnamefont {Liu}}, \bibinfo
  {author} {\bibfnamefont {Z.}~\bibnamefont {Gao}}, \bibinfo {author}
  {\bibfnamefont {H.}~\bibnamefont {Lu}},\ and\ \bibinfo {author}
  {\bibfnamefont {X.~C.}\ \bibnamefont {Xie}},\ }\bibfield  {journal} {\bibinfo
   {journal} {Advanced Science}\ }\href
  {https://doi.org/10.1002/advs.202514818} {10.1002/advs.202514818} (\bibinfo
  {year} {2025})\BibitemShut {NoStop}%
\bibitem [{\citenamefont {Yang}\ \emph {et~al.}(2020)\citenamefont {Yang},
  \citenamefont {Gao}, \citenamefont {Liang},\ and\ \citenamefont
  {Wang}}]{Yang:2020mtz}%
  \BibitemOpen
  \bibfield  {author} {\bibinfo {author} {\bibfnamefont {S.-Z.}\ \bibnamefont
  {Yang}}, \bibinfo {author} {\bibfnamefont {J.-H.}\ \bibnamefont {Gao}},
  \bibinfo {author} {\bibfnamefont {Z.-T.}\ \bibnamefont {Liang}},\ and\
  \bibinfo {author} {\bibfnamefont {Q.}~\bibnamefont {Wang}},\ }\href
  {https://doi.org/10.1103/PhysRevD.102.116024} {\bibfield  {journal} {\bibinfo
   {journal} {Phys. Rev. D}\ }\textbf {\bibinfo {volume} {102}},\ \bibinfo
  {pages} {116024} (\bibinfo {year} {2020})},\ \Eprint
  {https://arxiv.org/abs/2003.04517} {arXiv:2003.04517 [hep-ph]} \BibitemShut
  {NoStop}%
\bibitem [{\citenamefont {Mameda}(2023)}]{Mameda:2023ueq}%
  \BibitemOpen
  \bibfield  {author} {\bibinfo {author} {\bibfnamefont {K.}~\bibnamefont
  {Mameda}},\ }\href {https://doi.org/10.1103/PhysRevD.108.016001} {\bibfield
  {journal} {\bibinfo  {journal} {Phys. Rev. D}\ }\textbf {\bibinfo {volume}
  {108}},\ \bibinfo {pages} {016001} (\bibinfo {year} {2023})},\ \Eprint
  {https://arxiv.org/abs/2305.02134} {arXiv:2305.02134 [hep-th]} \BibitemShut
  {NoStop}%
\bibitem [{\citenamefont {Yang}\ \emph {et~al.}(2025)\citenamefont {Yang},
  \citenamefont {Gao},\ and\ \citenamefont {Pu}}]{Yang:2024sfp}%
  \BibitemOpen
  \bibfield  {author} {\bibinfo {author} {\bibfnamefont {S.-Z.}\ \bibnamefont
  {Yang}}, \bibinfo {author} {\bibfnamefont {J.-H.}\ \bibnamefont {Gao}},\ and\
  \bibinfo {author} {\bibfnamefont {S.}~\bibnamefont {Pu}},\ }\href
  {https://doi.org/10.1103/PhysRevD.111.036013} {\bibfield  {journal} {\bibinfo
   {journal} {Phys. Rev. D}\ }\textbf {\bibinfo {volume} {111}},\ \bibinfo
  {pages} {036013} (\bibinfo {year} {2025})},\ \Eprint
  {https://arxiv.org/abs/2409.00456} {arXiv:2409.00456 [hep-ph]} \BibitemShut
  {NoStop}%
\bibitem [{Note4()}]{Note4}%
  \BibitemOpen
  \bibinfo {note} {We note that the covariant current $\protect \tilde {j}^\mu
  $, derived in Refs.~\cite {Yang:2020mtz,Mameda:2023ueq,Yang:2024sfp}, is
  defined in the inertial frame. Hence, it is related to our results through
  $\protect \tilde {j}^0 = n^{(EE)}$ and $\protect \tilde {\protect \bm
  {j}}=n^{(EE)}{\protect \bm {u}}+ {\protect \bm {j}}^{(EE)}$.}\BibitemShut
  {Stop}%
\bibitem [{\citenamefont {'t~Hooft}(1981)}]{tHooft:1981bkw}%
  \BibitemOpen
  \bibfield  {author} {\bibinfo {author} {\bibfnamefont {G.}~\bibnamefont
  {'t~Hooft}},\ }\href {https://doi.org/10.1016/0550-3213(81)90442-9}
  {\bibfield  {journal} {\bibinfo  {journal} {Nucl. Phys. B}\ }\textbf
  {\bibinfo {volume} {190}},\ \bibinfo {pages} {455} (\bibinfo {year}
  {1981})}\BibitemShut {NoStop}%
\bibitem [{\citenamefont {Luttinger}\ and\ \citenamefont
  {Kohn}(1955)}]{Luttinger-Kohn}%
  \BibitemOpen
  \bibfield  {author} {\bibinfo {author} {\bibfnamefont {J.~M.}\ \bibnamefont
  {Luttinger}}\ and\ \bibinfo {author} {\bibfnamefont {W.}~\bibnamefont
  {Kohn}},\ }\href {https://doi.org/10.1103/PhysRev.97.869} {\bibfield
  {journal} {\bibinfo  {journal} {Phys. Rev.}\ }\textbf {\bibinfo {volume}
  {97}},\ \bibinfo {pages} {869} (\bibinfo {year} {1955})}\BibitemShut
  {NoStop}%
\bibitem [{\citenamefont {Schrieffer}\ and\ \citenamefont
  {Wolff}(1966)}]{Schrieffer-Wolff}%
  \BibitemOpen
  \bibfield  {author} {\bibinfo {author} {\bibfnamefont {J.~R.}\ \bibnamefont
  {Schrieffer}}\ and\ \bibinfo {author} {\bibfnamefont {P.~A.}\ \bibnamefont
  {Wolff}},\ }\href {https://doi.org/10.1103/PhysRev.149.491} {\bibfield
  {journal} {\bibinfo  {journal} {Phys. Rev.}\ }\textbf {\bibinfo {volume}
  {149}},\ \bibinfo {pages} {491} (\bibinfo {year} {1966})}\BibitemShut
  {NoStop}%
\bibitem [{\citenamefont {Kozii}\ \emph {et~al.}(2021)\citenamefont {Kozii},
  \citenamefont {Avdoshkin}, \citenamefont {Zhong},\ and\ \citenamefont
  {Moore}}]{Kozii}%
  \BibitemOpen
  \bibfield  {author} {\bibinfo {author} {\bibfnamefont {V.}~\bibnamefont
  {Kozii}}, \bibinfo {author} {\bibfnamefont {A.}~\bibnamefont {Avdoshkin}},
  \bibinfo {author} {\bibfnamefont {S.}~\bibnamefont {Zhong}},\ and\ \bibinfo
  {author} {\bibfnamefont {J.~E.}\ \bibnamefont {Moore}},\ }\href
  {https://doi.org/10.1103/PhysRevLett.126.156602} {\bibfield  {journal}
  {\bibinfo  {journal} {Phys. Rev. Lett.}\ }\textbf {\bibinfo {volume} {126}},\
  \bibinfo {pages} {156602} (\bibinfo {year} {2021})}\BibitemShut {NoStop}%
\bibitem [{\citenamefont {Avdoshkin}\ and\ \citenamefont
  {Popov}(2023)}]{Avdoshkin}%
  \BibitemOpen
  \bibfield  {author} {\bibinfo {author} {\bibfnamefont {A.}~\bibnamefont
  {Avdoshkin}}\ and\ \bibinfo {author} {\bibfnamefont {F.~K.}\ \bibnamefont
  {Popov}},\ }\href {https://doi.org/10.1103/PhysRevB.107.245136} {\bibfield
  {journal} {\bibinfo  {journal} {Phys. Rev. B}\ }\textbf {\bibinfo {volume}
  {107}},\ \bibinfo {pages} {245136} (\bibinfo {year} {2023})}\BibitemShut
  {NoStop}%
\bibitem [{\citenamefont {Avdoshkin}(2024)}]{Avdoshkin:2024rch}%
  \BibitemOpen
  \bibfield  {author} {\bibinfo {author} {\bibfnamefont {A.}~\bibnamefont
  {Avdoshkin}},\ }\href@noop {} {\bibinfo {title} {{Geometry of degenerate
  quantum states, configurations of $m$-planes and invariants on complex
  Grassmannians}}} (\bibinfo {year} {2024}),\ \Eprint
  {https://arxiv.org/abs/2404.03234} {arXiv:2404.03234 [quant-ph]} \BibitemShut
  {NoStop}%
\bibitem [{\citenamefont {Ahn}\ \emph {et~al.}(2021)\citenamefont {Ahn},
  \citenamefont {Guo}, \citenamefont {Nagaosa},\ and\ \citenamefont
  {Vishwanath}}]{Ahn_2021}%
  \BibitemOpen
  \bibfield  {author} {\bibinfo {author} {\bibfnamefont {J.}~\bibnamefont
  {Ahn}}, \bibinfo {author} {\bibfnamefont {G.-Y.}\ \bibnamefont {Guo}},
  \bibinfo {author} {\bibfnamefont {N.}~\bibnamefont {Nagaosa}},\ and\ \bibinfo
  {author} {\bibfnamefont {A.}~\bibnamefont {Vishwanath}},\ }\href
  {https://doi.org/10.1038/s41567-021-01465-z} {\bibfield  {journal} {\bibinfo
  {journal} {Nature Physics}\ }\textbf {\bibinfo {volume} {18}},\ \bibinfo
  {pages} {290–295} (\bibinfo {year} {2021})}\BibitemShut {NoStop}%
\bibitem [{\citenamefont {Avdoshkin}\ \emph {et~al.}(2025)\citenamefont
  {Avdoshkin}, \citenamefont {Mitscherling},\ and\ \citenamefont
  {Moore}}]{Avdoshkin_prl}%
  \BibitemOpen
  \bibfield  {author} {\bibinfo {author} {\bibfnamefont {A.}~\bibnamefont
  {Avdoshkin}}, \bibinfo {author} {\bibfnamefont {J.}~\bibnamefont
  {Mitscherling}},\ and\ \bibinfo {author} {\bibfnamefont {J.~E.}\ \bibnamefont
  {Moore}},\ }\href {https://doi.org/10.1103/w761-8nf7} {\bibfield  {journal}
  {\bibinfo  {journal} {Phys. Rev. Lett.}\ }\textbf {\bibinfo {volume} {135}},\
  \bibinfo {pages} {066901} (\bibinfo {year} {2025})}\BibitemShut {NoStop}%
\bibitem [{\citenamefont {Mitscherling}\ \emph {et~al.}(2025)\citenamefont
  {Mitscherling}, \citenamefont {Avdoshkin},\ and\ \citenamefont
  {Moore}}]{Mitscherling:2024uxz}%
  \BibitemOpen
  \bibfield  {author} {\bibinfo {author} {\bibfnamefont {J.}~\bibnamefont
  {Mitscherling}}, \bibinfo {author} {\bibfnamefont {A.}~\bibnamefont
  {Avdoshkin}},\ and\ \bibinfo {author} {\bibfnamefont {J.~E.}\ \bibnamefont
  {Moore}},\ }\href {https://doi.org/10.1103/qscv-qxqt} {\bibfield  {journal}
  {\bibinfo  {journal} {Phys. Rev. B}\ }\textbf {\bibinfo {volume} {112}},\
  \bibinfo {pages} {085104} (\bibinfo {year} {2025})},\ \Eprint
  {https://arxiv.org/abs/2412.03637} {arXiv:2412.03637 [cond-mat.str-el]}
  \BibitemShut {NoStop}%
\bibitem [{\citenamefont {Het{\'e}nyi}\ and\ \citenamefont
  {L{\'e}vay}(2023)}]{Hetenyi:2023hye}%
  \BibitemOpen
  \bibfield  {author} {\bibinfo {author} {\bibfnamefont {B.}~\bibnamefont
  {Het{\'e}nyi}}\ and\ \bibinfo {author} {\bibfnamefont {P.}~\bibnamefont
  {L{\'e}vay}},\ }\href {https://doi.org/10.1103/PhysRevA.108.032218}
  {\bibfield  {journal} {\bibinfo  {journal} {Phys. Rev. A}\ }\textbf {\bibinfo
  {volume} {108}},\ \bibinfo {pages} {032218} (\bibinfo {year} {2023})},\
  \Eprint {https://arxiv.org/abs/2309.03621} {arXiv:2309.03621 [quant-ph]}
  \BibitemShut {NoStop}%
\bibitem [{\citenamefont {Akamatsu}\ and\ \citenamefont
  {Yamamoto}(2013)}]{Akamatsu:2013pjd}%
  \BibitemOpen
  \bibfield  {author} {\bibinfo {author} {\bibfnamefont {Y.}~\bibnamefont
  {Akamatsu}}\ and\ \bibinfo {author} {\bibfnamefont {N.}~\bibnamefont
  {Yamamoto}},\ }\href {https://doi.org/10.1103/PhysRevLett.111.052002}
  {\bibfield  {journal} {\bibinfo  {journal} {Phys. Rev. Lett.}\ }\textbf
  {\bibinfo {volume} {111}},\ \bibinfo {pages} {052002} (\bibinfo {year}
  {2013})},\ \Eprint {https://arxiv.org/abs/1302.2125} {arXiv:1302.2125
  [nucl-th]} \BibitemShut {NoStop}%
\bibitem [{\citenamefont {Chen}\ \emph {et~al.}(2014)\citenamefont {Chen},
  \citenamefont {Son}, \citenamefont {Stephanov}, \citenamefont {Yee},\ and\
  \citenamefont {Yin}}]{Chen:2014cla}%
  \BibitemOpen
  \bibfield  {author} {\bibinfo {author} {\bibfnamefont {J.-Y.}\ \bibnamefont
  {Chen}}, \bibinfo {author} {\bibfnamefont {D.~T.}\ \bibnamefont {Son}},
  \bibinfo {author} {\bibfnamefont {M.~A.}\ \bibnamefont {Stephanov}}, \bibinfo
  {author} {\bibfnamefont {H.-U.}\ \bibnamefont {Yee}},\ and\ \bibinfo {author}
  {\bibfnamefont {Y.}~\bibnamefont {Yin}},\ }\href
  {https://doi.org/10.1103/PhysRevLett.113.182302} {\bibfield  {journal}
  {\bibinfo  {journal} {Phys. Rev. Lett.}\ }\textbf {\bibinfo {volume} {113}},\
  \bibinfo {pages} {182302} (\bibinfo {year} {2014})},\ \Eprint
  {https://arxiv.org/abs/1404.5963} {arXiv:1404.5963 [hep-th]} \BibitemShut
  {NoStop}%
\bibitem [{\citenamefont {Chen}\ \emph {et~al.}(2015)\citenamefont {Chen},
  \citenamefont {Son},\ and\ \citenamefont {Stephanov}}]{Chen:2015gta}%
  \BibitemOpen
  \bibfield  {author} {\bibinfo {author} {\bibfnamefont {J.-Y.}\ \bibnamefont
  {Chen}}, \bibinfo {author} {\bibfnamefont {D.~T.}\ \bibnamefont {Son}},\ and\
  \bibinfo {author} {\bibfnamefont {M.~A.}\ \bibnamefont {Stephanov}},\ }\href
  {https://doi.org/10.1103/PhysRevLett.115.021601} {\bibfield  {journal}
  {\bibinfo  {journal} {Phys. Rev. Lett.}\ }\textbf {\bibinfo {volume} {115}},\
  \bibinfo {pages} {021601} (\bibinfo {year} {2015})},\ \Eprint
  {https://arxiv.org/abs/1502.06966} {arXiv:1502.06966 [hep-th]} \BibitemShut
  {NoStop}%
\bibitem [{\citenamefont {Hayata}\ \emph {et~al.}(2021)\citenamefont {Hayata},
  \citenamefont {Hidaka},\ and\ \citenamefont {Mameda}}]{Hayata2020}%
  \BibitemOpen
  \bibfield  {author} {\bibinfo {author} {\bibfnamefont {T.}~\bibnamefont
  {Hayata}}, \bibinfo {author} {\bibfnamefont {Y.}~\bibnamefont {Hidaka}},\
  and\ \bibinfo {author} {\bibfnamefont {K.}~\bibnamefont {Mameda}},\ }\href
  {https://doi.org/10.1007/JHEP05(2021)023} {\bibfield  {journal} {\bibinfo
  {journal} {J. High Energy Phys.}\ }\textbf {\bibinfo {volume} {05}},\
  \bibinfo {pages} {023}},\ \Eprint {https://arxiv.org/abs/2012.12494}
  {arXiv:2012.12494 [hep-th]} \BibitemShut {NoStop}%
\bibitem [{\citenamefont {Liu}\ \emph {et~al.}(2019)\citenamefont {Liu},
  \citenamefont {Gao}, \citenamefont {Mameda},\ and\ \citenamefont
  {Huang}}]{Liu:2018xip}%
  \BibitemOpen
  \bibfield  {author} {\bibinfo {author} {\bibfnamefont {Y.-C.}\ \bibnamefont
  {Liu}}, \bibinfo {author} {\bibfnamefont {L.-L.}\ \bibnamefont {Gao}},
  \bibinfo {author} {\bibfnamefont {K.}~\bibnamefont {Mameda}},\ and\ \bibinfo
  {author} {\bibfnamefont {X.-G.}\ \bibnamefont {Huang}},\ }\href
  {https://doi.org/10.1103/PhysRevD.99.085014} {\bibfield  {journal} {\bibinfo
  {journal} {Phys. Rev. D}\ }\textbf {\bibinfo {volume} {99}},\ \bibinfo
  {pages} {085014} (\bibinfo {year} {2019})},\ \Eprint
  {https://arxiv.org/abs/1812.10127} {arXiv:1812.10127 [hep-th]} \BibitemShut
  {NoStop}%
\end{thebibliography}%

\end{document}